\documentstyle[11pt,aaspp4,epsfig]{aastex}

\begin{document}

\title{Early Tracking Behavior in Small-field Quintessence Models}

\author{Wei Wang$^1$, Bo Feng$^2$}
\affil{$^1$National Astronomical Observatories, Chinese Academy of Sciences, Beijing  100012, China, wwang@lamost.bao.ac.cn \\
$^2$Institute of High Energy Physics, Chinese Academy of Sciences,
P.O. Box 918-4, Beijing 100049, China, fengbo@mail.ihep.ac.cn}


\begin{abstract}
We study several quintessence models which are singular at $Q=0$,
and use a simple initial constraint $Q_i\ge H_{inflation}/2\pi$ to
see when they enter tracking regime, disregarding the details of
inflation. We find it can give strong constraints for the inverse
power-law potential $V=V_0Q^{-\alpha}$, which has to enter tracking
regime for ${\rm ln}z \sim 10$. While for the supergravity model
$V=V_0Q^{-\alpha}{\rm exp}(kQ^2/2)$, the constraint is much
weakened. For another kind inverse power-law potential $V=V_0{\rm
exp}(\lambda/Q)$, it exhibits no constraints.

\end{abstract}


\section{Introduction}
Recent observations by type Ia supernova(SN) survey(Garnavich et
al. 1998, Perlmutter et al. 1998) and cosmic microwave
background(CMB) anisotropies(de bernardis et al. 2002, Lee et al.
2001) strongly show the evidence for a cosmological constant or
dark energy. In general case, the dark energy (or Quintessence)
can have a time-dependent equation of state,
$P_Q=\omega_Q(t)\nu_Q$ (Wetterich 1988, Peebles \& Ratra 1988),
which is invoked to explain the coincidence problem. Some
researchers have reanalyzed the cosmological data from CMB, SN,
large scale structure(LSS) and gravitational lens statistics(Bean
\& Melchiorri 2001, Baccigalupi et al. 2001, Hannestad \&
M\"ortsell, 2002, Chae et al. 2002), confirming that quintessence
is slightly preferred with respect to cosmological constant.

An important class of quintessence models are known as tracking
models. By coupling a scalar field to matter one can obtain
tracking solutions ( Zlatev et al. 1998, Steinhardt et al. 1999)
for time dependence of dark energy density so that it always
follows the dominant energy density component and fairly
independent of initial conditions. Recently Malquarti \&
Liddle(2002) used stochastic inflation formalism to constrain the
initial values of quintessence after inflation. They have shown
that for inverse power-law form of quintessence $V=V_0Q^{-\alpha}$
satisfying current observations, initial $Q$ was so large that it
could not enter tracking regime until the matter-domination epoch.
This has put the tracking behavior in considerable jeopardy for
such quintessence models.

In the present paper, we will study the behaviors of several
quintessence models which are singular at $Q=0$(dubbed
"Small-field Quintessence Models") affected by inflation.
Quintessence being almost massless, one has $\delta Q \sim H/2\pi$
(See Liddle et al 1993 for details ) after inflation. The simplest
constraint on initial $Q$ is that its value should be larger than
the perturbative part: $Q \ge \delta Q$, i.e. $Q\ge H/2\pi$. We
shall use this constraint to see when they enter tracking regime,
disregarding the details of inflation. In the following section,
we will discuss the tracking behaviors of three quintessence
models: inverse power-law potential $V=V_0Q^{-\alpha}$ (Ratra \&
Peebles 1988), the supergravity model $V=V_0Q^{-\alpha}{\rm
exp}(kQ^2/2)$ and the exponential form of inverse power-law
potential $V=V_0{\rm exp}(\lambda/Q)$.

\section{Models and tracking solutions}
We shall consider models of quintessence in a flat cosmological
background, i.e. $\Omega_k=0$. The ratio of energy density to the
critical density today is $\Omega_Q$ for the $Q$-field and
$\Omega_m$ for the matter density where $\Omega_m + \Omega_Q=1$.
We also define a background equation-of-state $\omega_B$,
$\omega_B=1/3$ in radiation-dominated epoch and 0 in
matter-dominated era. We use dimensionless units where the Planck
mass is M$_{pl}=$1.

The equation of motion for the $Q$-field is
\begin{equation}
\ddot{Q}+3H\dot{Q}+V'=0,
\end{equation}
where $V'=\ddot{Q}{1-\omega_Q \over 1+\omega_Q}$ and
\begin{equation}
H^2=({\dot{a}\over a})^2={8\pi G\over 3}(\rho_Q+\rho_B),
\end{equation}
$a$ is the Robertson-Walker scale factor, $\rho_B=\rho_m+\rho_r$,
$\rho_m$ and $\rho_r$ are the matter and radiation energy density
respectively. Early in the radiation-dominated epoch,
 we have $H \approx 1/{2t}$ and substituting the form of $V'$ into
 Eq.(1):
\begin{equation}
\ddot{Q}+{3(1+\omega_Q) \over 4t}\dot{Q}=0.
\end{equation}
Assuming $\omega_Q$ to be a constant, we can easily obtain the
solution of the equation:$\dot{Q}= C t^{-3(1+\omega_Q)/4}$, where
C is a constant. So we have
\begin{equation}
\Omega_Q=({\dot{Q} \over H})^2\propto t^{-{3\over 2}(1+\omega_Q)+2}\propto a^{-3(1+\omega_Q)+4}.
\end{equation}
The above equation gives the rough evolution of $Q$-field. On the
other hand we can obtain the $Q$ energy density in another method.
According to the present cosmological density $\rho_0$, we can
write
\begin{equation}
\rho_Q = \rho_0({a_{\rm eq}\over a_0})^{-3(1+\omega_1)}({a\over a_{\rm eq}})^{-3(1+\omega_2)},
\end{equation}
or
\begin{equation}
{\rm ln}\rho_Q = {\rm ln}\rho_0 + 3(\omega_1-\omega_2){\rm ln}(z_{\rm eq}+1)+3(1+\omega_2){\rm ln}(z+1),
\end{equation}
where $z$ is the redshift, the subscript eq denotes the epoch of
matter-radiation equality, and $\omega_1$ and $\omega_2$ are the
equations of state of Q-field during matter- and
radiation-domination epochs. Note in the analytical formula
Eqs.(5,6) we've assumed $\omega_1$ to be a constant and today is
matter-dominant, this has led to considerably uncertainties when
compared to the exact numerical case (See Fig.3).

An important function is $\Gamma\equiv V''V/(V')^2$, whose properties determine whether tracking solutions exist.
Taking the derivative of the equation-of-motion with respect to $Q$ and combining with the equation-of-motion itself,
we can obtain the tracking equation:
\begin{equation}
\Gamma\equiv 1+{\omega_B-\omega_Q \over
2(1+\omega_Q)}-{1+\omega_B-2\omega_Q \over 2(1+\omega_Q)}{\dot
x\over 6+x} -{2\over (1+\omega_Q)}{\ddot{x}\over (6+\dot x)^2},
\end{equation}
where $x\equiv (1+\omega_Q)/(1-\omega_Q), \dot x\equiv d{\rm
ln}x/d{\rm ln}a$ and $\ddot{x}\equiv d^2{\rm ln}x/d{\rm ln}a^2$.

In the following sections, we will  discuss the tracking behavior of
different quintessence models in detail. And we take the
cosmological parameters derived from recent observational
constraints throughout, $\Omega_m=0.3, \omega_Q=-0.82$ and the
Hubble constant $h=0.65$.

\subsection{Pure Inverse power-law models}
The quintessence models of the pure inverse power-law potentials, as originally introduced by Ratra \& Peebles (1988):
$V=V_0Q^{-\alpha}$. For the tracking solution,
\begin{equation}
\Gamma-1={\omega_B-\omega_Q \over 2(1+\omega_Q)}={1\over \alpha},
\end{equation}
then $\omega_Q=\alpha\omega_Q-2/(\alpha+2)$. Since
$\omega_1=-2/(\alpha+2)$ $(\omega_B=0)$ and
$\omega_2=\alpha-6/(3\alpha+6)$  $(\omega_B=1/3)$, according to
Eq. (6), we have
\begin{equation}
{\rm ln}\rho_Q={\rm ln}\rho_0 - {\alpha\over \alpha+2}{\rm ln}(z_{\rm eq}+1)+{4\alpha+12\over \alpha+2}{\rm ln}(z+1).
\end{equation}

For $\omega_Q=\rho_Q-2V/\rho_Q$, we then have
$V_0Q^{-\alpha}=V=\rho_Q(1-\omega_Q)/2$, i.e.  ${\rm
ln}V_0-\alpha{\rm ln}Q= {\rm ln}{1-\omega_Q\over 2}\rho_Q$. Taking
the approximation $V_0\simeq \rho_0$, we can obtain the final
analytical form:
\begin{equation}
{\rm ln}Q=({\rm ln}\rho_0 - {\rm ln}{1-\omega_Q\over 2}\rho_Q)/\alpha.
\end{equation}

We have also computed the tracking behavior of the quintessence
models numerically according to Eqs. (1) and (2). In Fig. 1, the
dashed ($\alpha=1.4$) and dotted ($\alpha=0.67$) lines show the
evolution of $Q$-field with the different values of $\alpha$. We
have $Q \varpropto (z+1)*Exp(-(4\alpha+12)/\alpha=/(\alpha=+2))$
and $Q\sim 1$ today. For a smaller $\alpha$, $Q$ would be smaller
earlier in the tracking regime, as shown in Fig.1. Our fit with
$\Omega_m=0.3, \omega_Q=-0.82$ requires $\alpha \approx 0.67$ and
$V_0^{1/4}=1.8* 10^{-31}$. In Malquarti \& Liddle(2002), their
fitting gives $0\le\alpha\le 1$(68\% confidence), our result is in
agreement with theirs.

\subsection{Supergravity models}
In this section we consider the supergravity version of model
considered previously with a superpotential of the form $V\propto
Q^{-\alpha}$. We take the potential form $V=V_0Q^{-\alpha}{\rm
exp}(kQ^2/2)$ (Brax et al 2000), where $k=8 \Pi G$, and study its
tracking behavior.

For tracking solutions, we have
\begin{equation}
\Gamma-1={(k+\alpha Q^{-2})V^2 + (kQ-\alpha Q^{-1})^2 V^2 \over
(kQ-\alpha Q^{-1})^2 V^2}-1={k+\alpha Q^{-2}
 \over (kQ-\alpha Q^{-1})^2}.
\end{equation}
Because $k\ll \alpha Q^{-2}(Q\to 0)$, $\Gamma-1 \simeq 1/\alpha$.
It is similar with former
 inverse power-law potentials. For $\omega_Q=-0.82$, and $\Omega_m=0.3$, $\alpha=11$  and
$V_0^{1/4}=1.93 \times 10^{-32}$ are expected.

In Fig. 1, the solid line shows the $Q$-field evolution with the
redshift. And in Fig. 2, we also show
 the evolution of the equation-of-state $\omega_Q$ in both the inverse power-law  and the supergravity potentials,
 taking the same parameter $\alpha=11$. In the early radiation-dominant epoch, they have the same equation of state,
 while later when matter and quintessence dominate, $Q$ grows larger and the factor $\exp(kQ^2/2)$
 takes effect today in the supergravity  model, which makes the main contribution to the different behaviors of the two models
 today. The pure inverse power law form $V=V_0 Q^{-11}$ is ruled out for it predicts $\omega_Q >-0.3$ and cannot give an
  accelerating universe today, meanwhile  $V=V_0 Q^{-\alpha}{\rm
exp}(kQ^2/2)$ is still not excluded. In the two models, we have
given the analytical forms for the tracking behavior according the
Eqs. (9) and (10); we have also compared the results of analytical
and numerical calculations in the supergravity version in Fig.3.

\subsection{Exponential Form of inverse power-law model}
Here, we further consider another kind of inverse power-law
potential with the form like $V=V_0e^{\lambda/Q}$. In our case,
$\lambda \approx 0.3$ and $V_0^{1/4}=1.85 \times 10^{-31}$. However,
since the equation-of-state of the model $\omega_Q$ varies with $t$,
it is relatively difficult to obtain the analytical solutions.
 In Fig. 4 we have shown the evolution of the $Q$-field and the equation-of-state $\omega_Q$ by numerical calculations,
 the tracking behavior is different from the previous two models discussed above. The evolution curve of $Q$
 -field always rises with time. We are able to give some rough estimations since $Q<<1$ is also satisfied for large
 $z$. The form $e^{\lambda/Q}$ can be expanded to $Q^{-\alpha}$ series with $\alpha \rightarrow \infty $ ,
 hence $\omega_Q=(\alpha-6)/(3\alpha+6)  \approx 1/3$. Its early
 behavior of $Q$ and $\rho_Q$ can also be explained as $\alpha \rightarrow \infty
 $ , where $Q$ would be much larger than in the pure inverse power-law models from above analysis.

For the comparison of the three models, we consider the very early
behavior of $Q$-field. In the very early time, the three models take
the same form $V=V_0 Q^{-\alpha}$ , with $\alpha=0.67$, $11$ and
$\infty$, respectively.

In Fig. 5, we show our constraints on the tracking behavior of the
quintessence models by taking the initial condition $Q_i=H/{2\pi}$,
where $H\sim 10^{-5}$. We take $\dot{Q_i}=0$ initially. We do not
include the exponential form of inverse power-law model because we
have no constraints on it, as can be seen from Fig.4. The solid line
displays the evolution of $Q$-field in the supergravity model, while
the dashed line denotes the inverse power-law model. For
$V=V_0Q^{-\alpha}$, it can only enter tracking regime after ${\rm
ln}z \sim 10$. For the supergravity model $V=V_0Q^{-\alpha}{\rm
exp}(kQ^2/2)$, it requires $\ln(z+1) \le 43$ .

\section{Conclusions and Discussions}
We have analyzed the dynamical evolution and tracking solutions of
three quintessence models. We used a simple constraint $Q_i\ge
H/{2\pi}$ to study the tracking behaviors, and found that it can
also give a strong constraint on the pure inverse power-law model
which enters tracking regime at a late stage ${\rm ln}z \sim 10$.
The key fact is that for such a pure inverse power-law model,
$\alpha$ has to be very small in order to fit current observations.
While for the supergravity model and exponential form of inverse
power-law model, the exponential form  takes a positive effect,
rendering them have $\omega_Q \sim -1$ today and satisfy the CMB and
SN constraints, meanwhile they take the pure inverse power-law form
with much larger $\alpha$ and hence little constraint is exhibited
with $Q_i\ge H/{2\pi}$ when entering the tracking regime.

The typical tracking behavior of inverse power-law model begins only
at quite a late stage of evolution, well after nucleosynthesis and
possibly after decoupling too as presented by this paper and
Malquarti \& Liddle. Tracking is the key to solve the coincidence
problem, tracking quintessence would lose the significance if it has
to enter tracking regime extremely late. Therefore, the supergravity
and exponential form of inverse power law model models show better
tracking behaviors in our analysis.

\acknowledgments We are grateful to Mingzhe Li, Xiulian Wang, De-Hai
Zhang, Xinmin Zhang and Yongheng Zhao for helpful discussions. This
work is supported by National Natural Science Foundation of China
under grant 10273011.

\begin{figure}
\psfig{figure=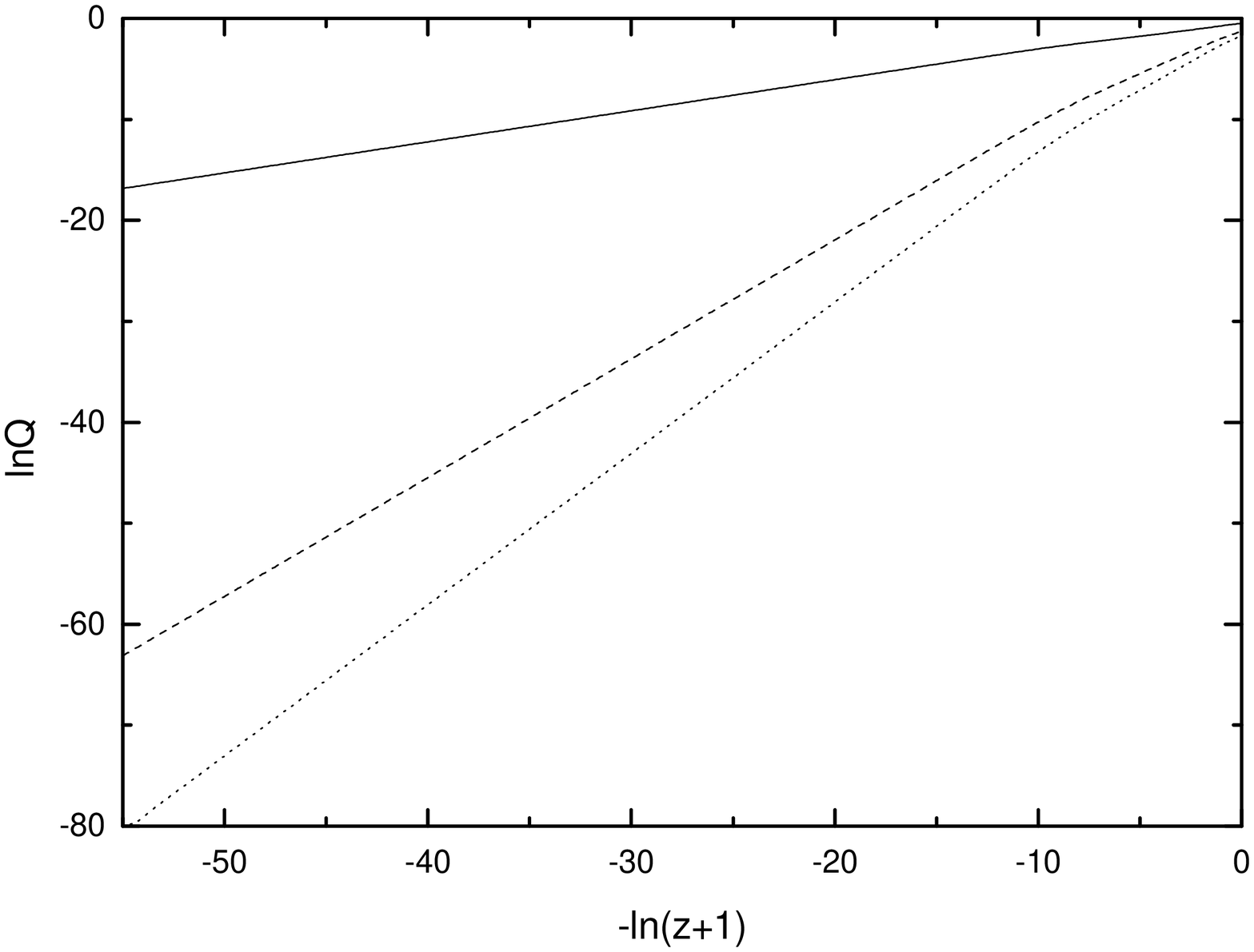,angle=0,width=16cm} \caption{The evolution of
the $Q$-field with the redshift. The solid line denotes the
supergravity model, and the dashed and dotted lines display inverse
power-law models with $\alpha=1.4$ and $\alpha=0.67$ respectively.}
\end{figure}

\begin{figure}
\psfig{figure=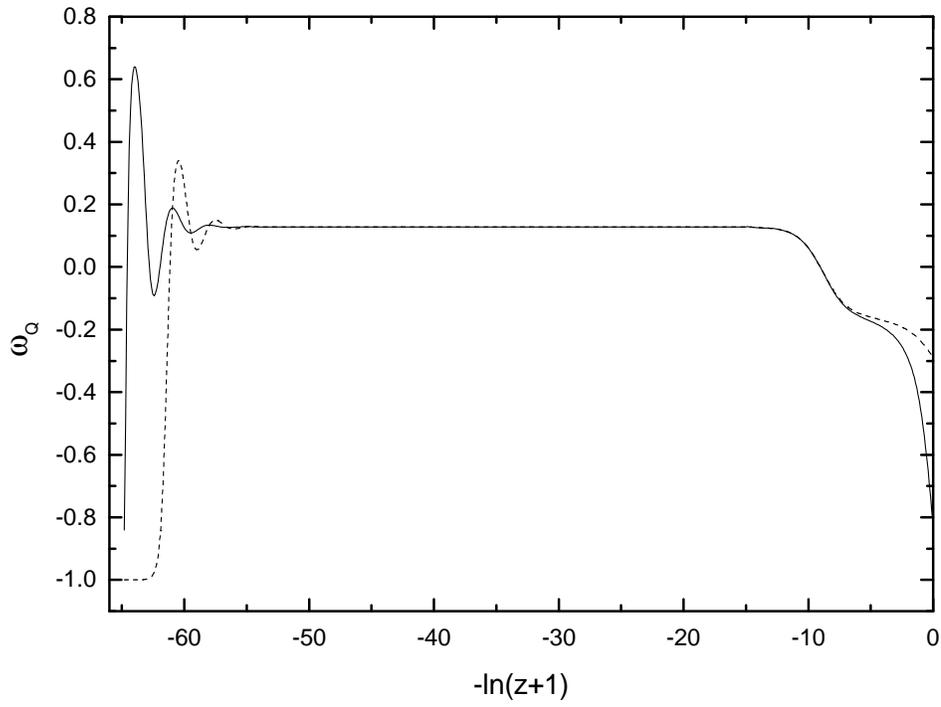,width=16cm} \caption{The different tracking
behaviors between the inverse power-law model(dashed) and the
supergravity model(solid), where $\alpha=11$.}
\end{figure}

\begin{figure}
\psfig{figure=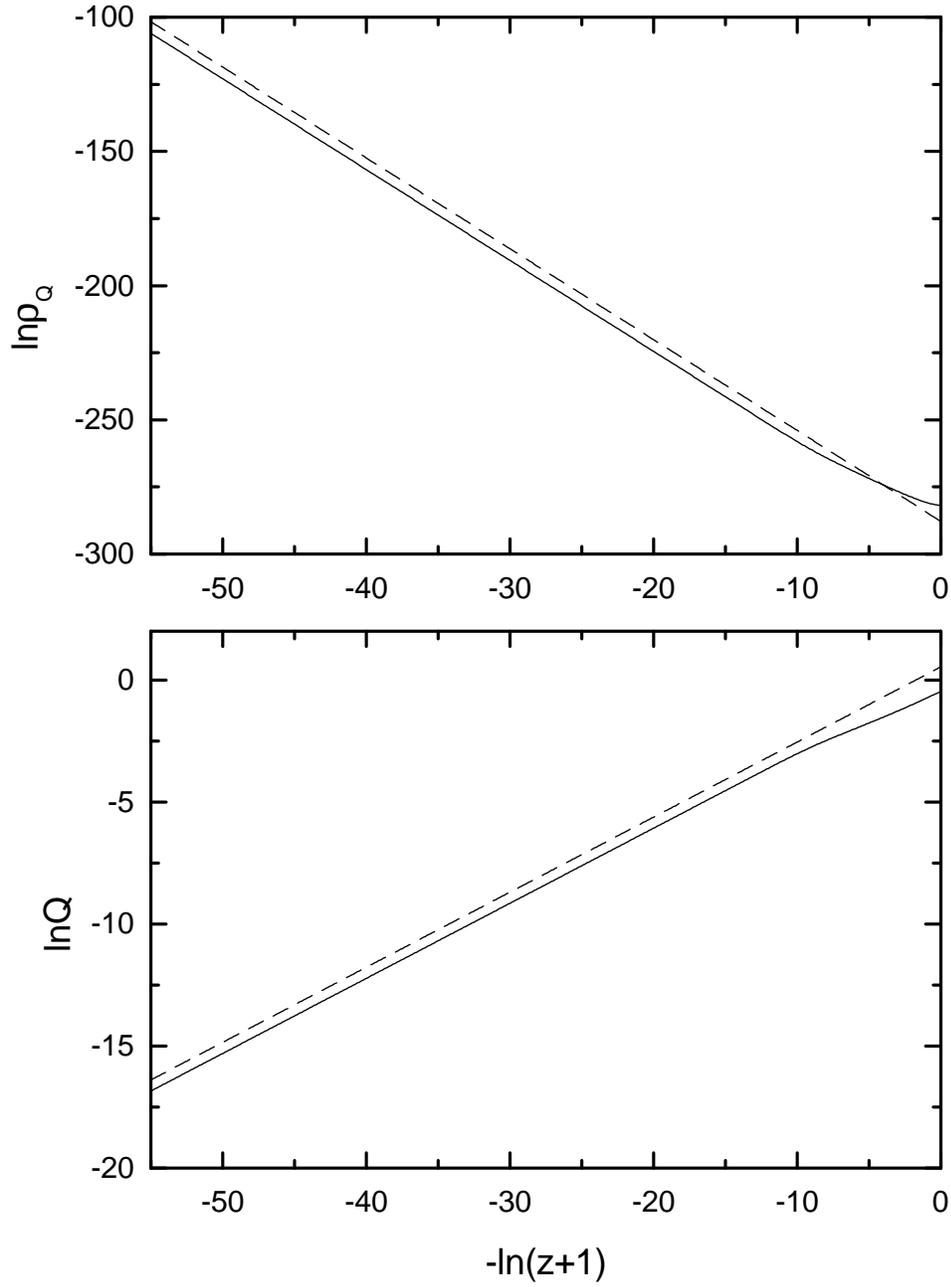,width=14cm} \caption{Comparison between the
analytical(solid) and numerical(dashed) tracking solutions in the
supergravity model.}
\end{figure}

\begin{figure}
\psfig{figure=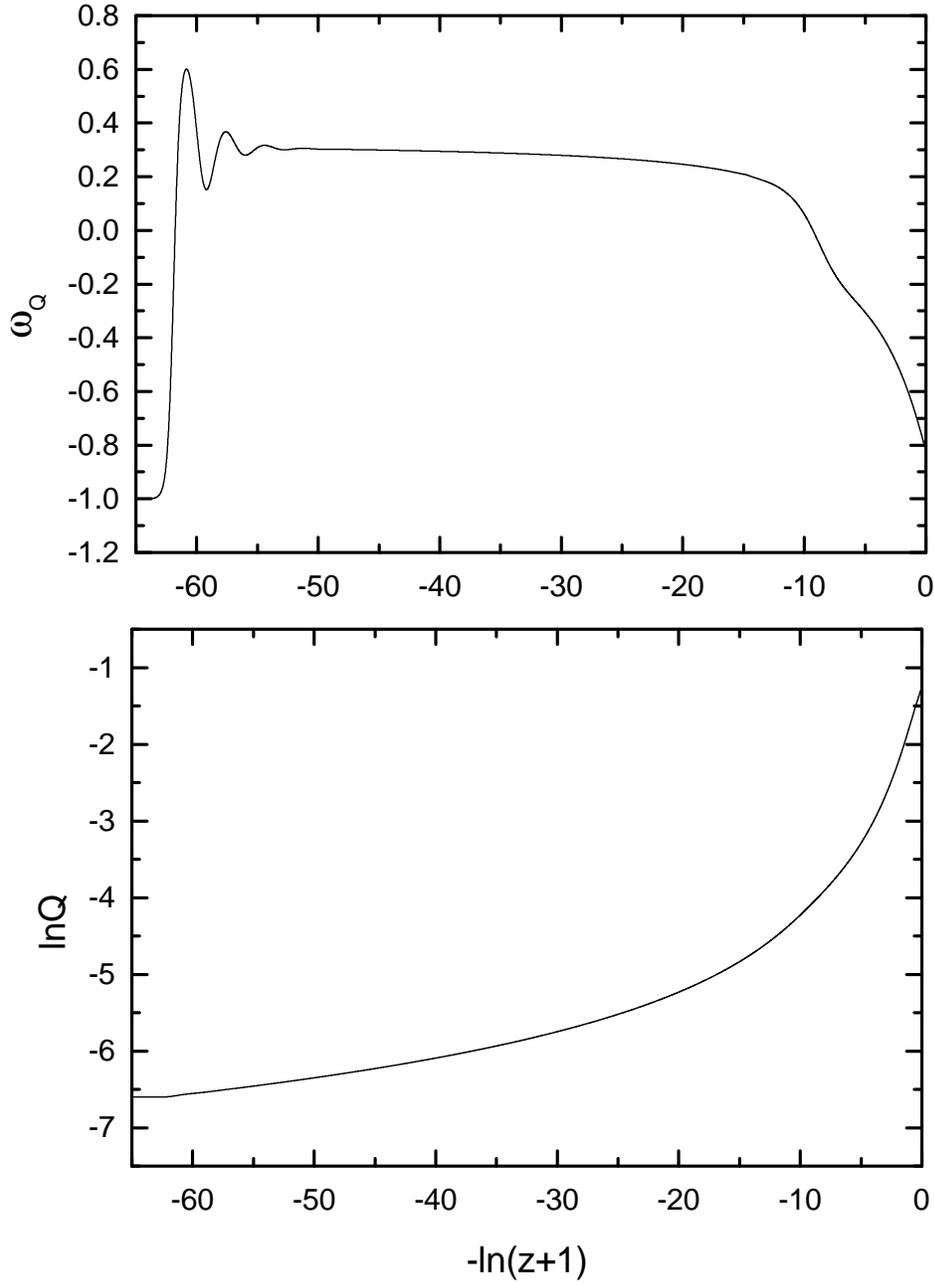,width=14cm} \caption{Evolution of the
equation-of-state $\omega_Q$ and $Q$-field with  redshift in the
exponential form of inverse power-law model. We found that no limit
on the tracking time can be obtained here. See the text for
details.}
\end{figure}

\begin{figure}
\psfig{figure=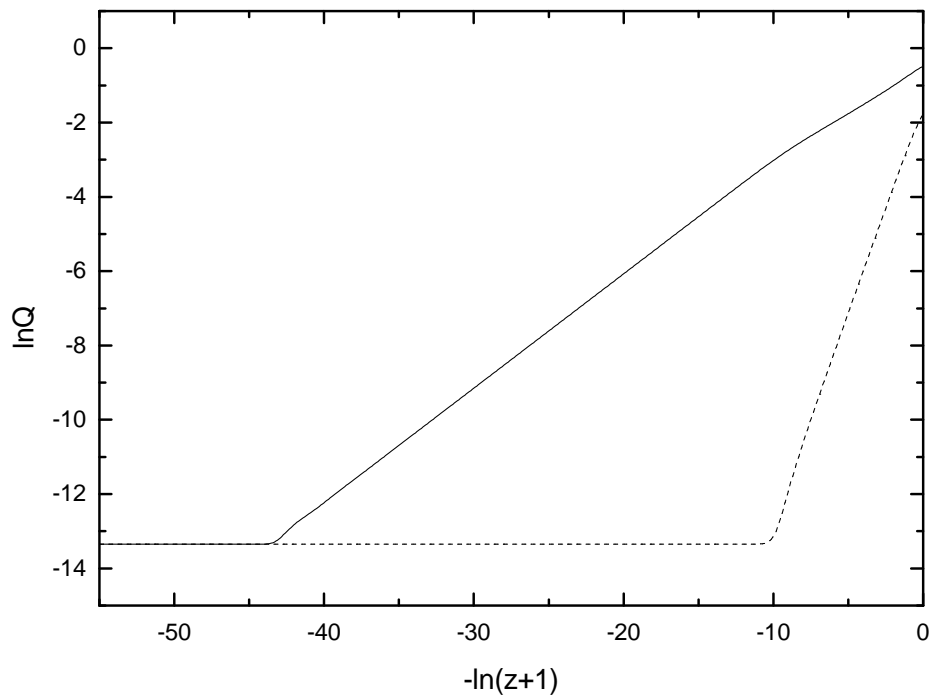,width=16cm} \caption{Evolution of the
$Q$-field with redshift in both inverse power-law(dashed) and
supergravity(solid) models. We take the initial condition
$Q_i=H/{2\pi}$, where $H\sim 10^{-5}$, and find that the time the
supergravity case enters the tracking regime is much earlier than
that of the inverse power-law model.}
\end{figure}

\end{document}